\documentclass[aps,prc,12pt,showpacs,showkeys,superscriptaddress,groupedaddress]{revtex4-1}

\usepackage[english]{babel}
\usepackage{indentfirst}
\usepackage{amssymb}
\usepackage{braket}
\usepackage{bm}
\usepackage{amsxtra}
\usepackage{amsmath}
\usepackage{supertabular}
\usepackage{multirow}
\usepackage{color}
\usepackage [mathcal]{eucal}
\usepackage{graphicx}
\usepackage{graphics}
\usepackage{lipsum}

\def\beq{\begin{equation}}
\def\eeq{\end{equation}}
\def\beqn{ \begin{eqnarray} }
\def\eeqn{ \end{eqnarray} }
\def\r{\mbox{\boldmath $r$}}

\def\p{\mbox{\boldmath $p$}}
\def\q{\mbox{\boldmath $q$}}
\def\s1s2{{ \boldsymbol{\sigma}(1) \cdot \boldsymbol{\sigma}(2) }}
\def\t1t2{{ \boldsymbol{\tau}(1) \cdot \boldsymbol{\tau}(2)  }}

\newcommand{\eep} { $(e,e^{\,\prime}p)$ } 
\newcommand{\ee} { $(e,e^{\,\prime})$ } 
\begin{document}
\noindent
\title{From \eep to Neutrino Scattering
}

\author{Carlotta Giusti}
\affiliation{Dipartimento di Fisica, 
Universit\`{a} degli Studi di Pavia and \\
INFN, Sezione di Pavia, via Bassi 6 I-27100 Pavia, Italy}
\date{\today}

\bigskip

\begin{abstract} 
Since a long time electron scattering has been envisaged as a powerful and preferential tool to investigate nuclear properties. In particular, the \eep knockout reaction has provided a wealth of information on the single particle (s.p.) aspects of nuclear structure, on the validity and the limit of the independent particle shell model (IPSM). The work done for electron scattering is extremely useful also for the analysis and the interpretation of neutrino oscillation experiments, where nuclei are used as neutrino detectors and it is crucial that nuclear effects in neutrino-nucleus interactions are well under control. In this contribution it is discussed if and how the work done for \eep can be exploited for the analysis of neutrino-nucleus scattering data.  
\end{abstract}

\bigskip
\bigskip
\bigskip


\maketitle

\section{Introduction}
\label{sec:intro}

Electron scattering is probably the best tool for investigating the
structure of atomic nuclei and their constituents.  The electromagnetic interaction is well known from quantum electrodynamic and it is weak 
compared with the strength of the interaction between hadrons. Thus, electron scattering is adequately treated assuming the validity of the Born approximation, i.e. the one-photon exchange mechanism between electron and target. Moreover, the ability to vary simultaneously and independently the energy $\omega$ and momentum $\q$ of the exchanged virtual photon
transferred to the nucleus makes it possible to map the nuclear response as a
function of its excitation energy with a spatial resolution that can be 
adjusted to the scale of the process we want to study.
Several decades of theoretical and experimental work on electron scattering have yielded a wealth of detailed information on nuclear structure and interaction mechanisms \cite{bof93,bof96,benh08}.

In a schematic representation of the nuclear response as a function of 
$\omega$ and $q$ different kinematic regions can be identified. A large broad peak occurs at about $\omega=q^2/2m$, where $m$ is the nucleon mass. Its position corresponds to the elastic peak in electron scattering by a free nucleon. It is quite natural to assume that a quasifree process is responsible for such a peak, with a nucleon emitted quasielastically. Coincidence \eep experiments in the quasielastic (QE) region, where the knocked-out proton is detected in coincidence with the scattering electron and the residual nucleus is left in any bound or continuum state, confirm such a picture and represent a valuable source of information on the single-nucleon degrees of freedom inside nuclei. Energy conservation defines the missing energy $E_{\mathrm m}$, i.e. the excitation energy of the residual nucleus with respect to the target ground state. At low energies in the $E_{\mathrm m}$ distribution, in  the discrete part of the spectrum, there are sharp isolated peaks corresponding to discrete eigenstates. Thus, for a value of $E_{\mathrm m}$ corresponding to a peak the residual nucleus is left in a discrete eigenstate: this is the case of the exclusive \eep reaction where the final nuclear state is completely determined. Many high-resolution \eep\ experiments over a wide range of nuclei, carried out at Saclay~\cite{bof93,bof96,fru84,mou76,ber82},
NIKHEF~\cite{bof93,bof96,dew90,lap93}, MAMI~\cite{blo95}, and Jefferson Lab~\cite{Gao00,malov00}, provided accurate information on the s.p. aspects of nuclear structure. From the study of the missing energy and momentum dependence of the experimental cross sections and from the comparison with the theoretical results it was possible to assign specific quantum numbers and spectroscopic factors to the peaks. 
Of course the \eep reaction can be investigated also for larger values of $E_{\mathrm m}$, in the continuum part of the spectrum.

If only the scattered electron is detected, in the QE region one nucleon is emitted but the final nuclear state is not determined, the cross section is integrated over $E_{\mathrm m}$ and includes all available final nuclear states, not only low-lying discrete states but also states in the continuum part of the spectrum. This is the inclusive QE \ee scattering.  

The exclusive \eep knockout reaction, for transitions to discrete low-lying states of the residual nucleus, represents a preferential tool to investigate proton-hole states in the target, the validity and the limit of the IPSM, of the mean-field approximation (MFA). From \eep data it was possible to see shell-model states,  but the fact that the spectroscopic factors, extracted for these states  from the comparison between theoretical and experimental results, are lower than predicted by the IPSM is a clear indication of the need to go beyond the MFA and to include correlations. 

It has always been  great challenge for nuclear physics to envisage experiments able to study correlations. Of particular interest are the short-range correlations (SRC) due to the short-range part of the NN interaction. 
Although their contribution to the spectroscopic factors is small, SRC can be investigated in \eep experiments at high missing energies~\cite{src1,src2}, in the continuum, beyond the threshold for the emission of a second nucleon, where, on the other hand, other processes and contributions involving more nucleons may come into play. More direct information on SRC can be obtained from electron-induced two-nucleon knockout~\cite{bof96,giu91,giu97,giu98,bar04,giu07a,giu07b}. Exclusive two-nucleon knockout reactions where the states of the residual nucleus can be separated, performed at NIKHEF~\cite{kest,zond,ond1,ond2,sta} and at MAMI~\cite{ros,mid1,mid2,mak}, have given clear signatures of SRC.

The work done for electron scattering can be extended to neutrino scattering. The two situations present many similar aspects and the extension of the formalism and of the models is straightforward. Neutrino scattering can be used to obtain additional and complementary information on nuclear properties.  Although very interesting, this is not, however, the main aim of most neutrino experiments, which are better aimed at determining neutrino properties. In neutrino experiments nuclei are used as neutrino detectors and a precise and reliable determination of neutrino properties requires that nuclear effects in neutrino-nucleus interactions are well under control. To this main, it is possible to exploit the work done for electron scattering, where sophisticated models and a large amount of  accurate data are available and more data can in principle be obtained from new experiments able to investigate specific effects. 
In spite of many similar aspects, electron and neutrino scattering represent two different situations and it is not guaranteed that a model able to describe electron scattering data will be able to describe neutrino scattering data with the same accuracy. Anyhow, the huge amount of work done on electron scattering, which provided a wealth of detailed information on nuclear properties, makes electron scattering the best available guide to determine the predictive power of a nuclear model. 
 
In the present contribution, starting from \eep, it is discussed  if and how the work done for  \eep   can be useful for the analysis of neutrino experiments.

\section{The \eep reaction}
\label{sec:eep}

In the exclusive \eep reaction, for a value of $E_{\mathrm m}$ corresponding to a peak in the missing-energy distribution, the  coincidence cross section contains the one-hole spectral density function (OHSDF), i.e.
\begin{equation}
S(\p_1,\p^{\prime}_1;E_{\mathrm m})=\langle \Psi_{\mathrm{i}} |
a^+_{\p^{\prime}_1} \delta(E_{\mathrm m} -H)a_{\p_1}| \Psi_{\mathrm{i}} 
\rangle,
\label{eq:HSF}
\end{equation}
which in its diagonal form ($\p_1=\p^{\prime}_1$) gives the joint probability
of removing from the target a nucleon, with momentum $\p_1$, leaving the
residual nucleus in a state with energy $E_{\mathrm m}$ with respect to the
target ground state. 

In the  inclusive \ee scattering, the integral of the spectral density over the whole energy spectrum gives the one-body density matrix (OBDM) 
$\rho(\p_1,\p^{\prime}_1)$, that in its diagonal form gives the nucleon 
momentum distribution $n(\p_1)$, i.e. the probability of finding in the target a nucleon with given momentum $\p_1$. 

In the one-photon exchange approximation the most general 
form of the \eep cross section~\cite{bof96}  involves the contraction 
between a lepton tensor $L_{\mu\nu}$, which basically contains lepton kinematics,  and a hadron tensor $W^{\mu\nu}$, whose components are given by bilinear combinations of the Fourier transforms of the transition matrix elements of the 
nuclear current operator between initial and final nuclear states, i.e.
\begin{equation}
J^{\mu}({\mbox{\boldmath $q$}}) = \int \langle\Psi_{\mathrm{f}} 
|\hat{J}^{\mu}({\mbox{\boldmath $r$}})|\Psi_{\mathrm{i}}\rangle
{\mathrm{e}}^{\,{\mathrm{i}}{\footnotesize {\mbox{\boldmath $q$}}}\cdot 
{\footnotesize {\mbox{\boldmath $r$}}}}
{\mathrm d}{\mbox{\boldmath $r$}}.     \label{eq:jm}
\end{equation} 

The usual description of the exclusive \eep reaction in the QE region~\cite{bof96,bof82} assumes the direct knockout (DKO) mechanism, which is related to the impulse approximation (IA), i.e. to the assumption that the electromagnetic probe interacts directly through a one-body current only with the quasifree ejectile nucleon. In addition, if the final-state interactions (FSI) between the outgoing nucleon and the residual nucleus are neglected and the plane-wave (PW) approximation is adopted for the outgoing proton wave function, in the plane-wave IA (PWIA), the \eep cross section is factorized into the product of a kinematical factor, the (off-shell) electron-proton cross section, and the diagonal OHSDF, i.e.
\begin{equation}
S(\p_{\mathrm m},E_{\mathrm m})= \sum_{\alpha} S_\alpha(E_{\mathrm m})
|\phi_{\alpha}(\p_{\mathrm m})|^2,
\label{eq:pwia}
\end{equation}
where the missing momentum $\p_{\mathrm m}$ is the recoil momentum of the 
residual nucleus. At each value of $E_{\mathrm m}$ the momentum dependence of 
the spectral function is given by the momentum distribution of the quasi-hole 
states $\alpha$ produced in the target nucleus at that energy and described by 
the normalized overlap functions (OF) $\phi_\alpha$ between the target ground 
state and the states of the residual nucleus. The spectroscopic 
factor (s.f.)  $S_\alpha$ is the norm of the OF and gives the probability that the quasi-hole state $\alpha$ is a pure hole-state in the target. In the IPSM $\phi_\alpha$ are the 
s.p. states of the model and $S_\alpha =1(0)$ for occupied (empty) states. In 
reality, the strength of a quasi-hole state is
fragmented over a set of s.p. states and $0 \leq S_\alpha < 1$. The
fragmentation of the strength is due to correlations and the s.f. can thus give a measurement of correlation effects.  

The PWIA is a simple and conceptually clear picture which is able to describe the main qualitative features of \eep cross sections, but the analysis of data requires  a more refined model where also FSI are taken into account and a distorted wave function is adopted for the outgoing proton. In the distorted-wave IA (DWIA) the cross section still contains the OF and the s.f. but factorization is destroyed and the OHSDF enters the cross section in its non diagonal form. 

In the DWIA the transition matrix elements in Eq.~(\ref{eq:jm}) are obtained in a one-body representation as~\cite{bof96,bof82}
\begin{equation}
J^{\mu}(\q) = \int 
\chi^{(-)*}_{E_{\mathrm m},\alpha}(\r_1)\hat{j}^{\mu}(\r,\r_1) \phi_{\alpha}(\r_1)\left[S_\alpha(E_{\mathrm m})\right]^{1/2}
{\mathrm{e}}^{\,{\mathrm{i}} {\footnotesize {\q}}\cdot
{\footnotesize {\r}}} {\mathrm d}\r {\mathrm d} \r_1 ,
\label{eq:dwia}
\end{equation}
where
\begin{equation}
\chi^{(-)}_{E_{\mathrm m},\alpha}(\r_1) =
\langle\Psi_\alpha(E_{\mathrm m})|a_{\r_1}| \Psi_{\mathrm f}\rangle
\label{eq:dw}
\end{equation}
is the s.p. distorted wave function of the ejectile and the OF
\begin{equation}
\left[S_\alpha(E_{\mathrm m})\right]^{1/2}\phi_{\alpha}(\r_1) =
\langle\Psi_\alpha(E_{\mathrm m})|a_{\r_1} |\Psi_{\mathrm i}\rangle
\label{eq:ovf}
\end{equation}
describes the residual nucleus as a hole state in the target. The s.f.
$S_\alpha(E_{\mathrm m})$ gives the
probability of removing from the target a nucleon at $\r_1$ leaving the 
residual nucleus in the state $\Psi_\alpha(E_{\mathrm m},\r_1)$.

The scattering state in Eq.~(\ref{eq:dw}) and the bound state 
$\phi_{\alpha}(\r_1)$ in Eq.~(\ref{eq:ovf}) are consistently derived in this 
model from an energy-dependent non-Hermitian optical-model Hamiltonian.
A fully consistent calculation would be, however, extremely difficult and in usual applications phenomenological ingredients are generally employed. Calculations have been performed within a nonrelativistic DWIA~\cite{giu87,giu88,giu11} or a relativistic RDWIA~\cite{pick85,jin92,udi93,udi96,udi99,kel97,meu01a,meu01b,meu02a,meu02b}. In RDWIA a relativistic nuclear current and relativistic Dirac spinors for bound and scattering states are used. The nucleon scattering state is eigenfunction of a phenomenological (nonrelativistic in DWIA or relativistic in RDWIA) optical potential (OP), determined through a fit to elastic proton-nucleus scattering data. For the OF there are also some calculations including correlations but, in general, phenomenological MF wave functions are adopted, Woods-Saxon (WS) or Hartree-Fock (HF) wave functions in DWIA and relativistic wave functions from a MFA in RDWIA.  The results of this model, in its nonrelativistic and relativistic version, are able to give an excellent description of \eep data in a wide range of nuclei and in different kinematics~\cite{bof96}. 

\begin{figure}
\begin{center}
\includegraphics[scale=0.5, angle=0]{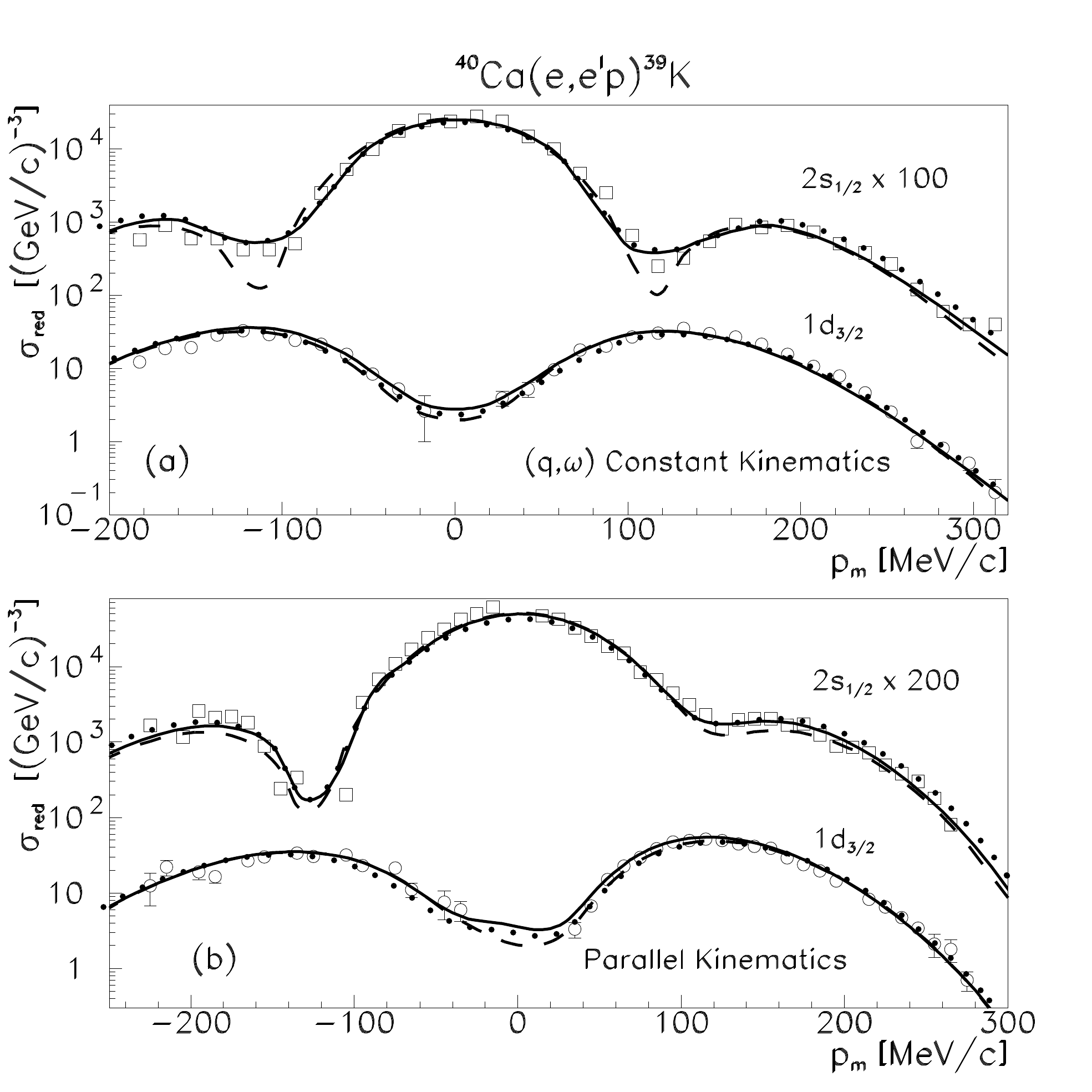} 
\caption{Reduced cross section of the $^{40}$Ca\eep reaction as a function of $p_{\mathrm m}$ for the transitions to the
$3/2^{+}$ ground state and to the $1/2^{+}$ excited state at 2.522 MeV of $^{39}$K, in 
($\q,\omega$) constant (upper panel) and parallel (lower panel) kinematics. 
For the calculations in ($\q,\omega$) constant kinematics the incident 
electron energy is $E_k= 483.2$ MeV, the electron scattering angle 
$61.52^{\circ}$, 
and $q=450$ MeV/$c$. In parallel kinematics $E_k= 440$ MeV. The outgoing proton energy is $100$ MeV in all the calculations. DWIA results with 
phenomenological WS (solid lines) and HF (dotted lines) wave functions and 
RDWIA (dashed lines) results are compared. Experimental data from Ref.~\cite{kra90t}. Positive (negative) values of $p_{\mathrm m}$ refer to situations 
where in ($\q,\omega$) constant kinematics 
the angle between the outgoing proton momentum $\p'$ and the incident electron 
$\p_k$ is larger (smaller) than the angle between $\q$ and $\p_k$, in parallel 
kinematics $|\q|<|\p'|$ ($|\q|>|\p'|$). (from Ref.~\cite{giu11}). 
}
\label{fig:ca40exp}
\end{center}
\end{figure}

Model calculations are able to give a good description of the shape of the experimental $p_{\mathrm m}$ distributions for transitions to discrete low-lying states of the residual nucleus. An example is shown in Fig.~\ref{fig:ca40exp} in comparison with the $^{40}$Ca\eep data measured at NIKHEF  for two different final states in parallel and ($\q,\omega$) constant kinematics \cite{kra90t}.
In the so-called parallel
kinematics \cite{bof96}, the momentum of the outgoing proton $\p'$ is
kept fixed and it is taken parallel, or antiparallel, to 
the momentum transfer $\q$.  Different values of  
$p_{\mathrm m}$ are obtained by varying the electron scattering angle
and, as a consequence, $q$.  In the so-called ($\q,\omega$) constant kinematics, $\q$ and $\p'$ are kept constant and the value of $p_{\mathrm m}$ is changed by varying the angle of the outgoing proton.
The quantity shown in the figure is the so-called reduced cross section,  i.e. the cross section divided by the kinematic factor and the free off-shell electron-proton cross section. This is the quantity that in PWIA gives the diagonal OHSDF of Eq.~(\ref{eq:pwia}), i.e. the momentum distribution of the quasi-hole state. The three results displayed in the figure are obtained in DWIA with two different WS and HF bound state wave functions and in RDWIA~\cite{giu11}. Coulomb distortion of the electron wave functions is included in the calculations. 
All the three theoretical results provide a good
description of the shape of the experimental distributions.  

In order to reproduce the magnitude of the experimental data, a
reduction factor must been applied to the
theoretical results. This factor is identified with the s.f. The so-called experimental s.f.'s obtained from this procedure give a depletion of the 
quasi-hole states near the Fermi energy of about 30-40\% with respect to the predictions of the IPSM~\cite{bof96} which can be attributed to NN correlations. The s.f.  gives a measurement of correlation effects, but since 
in these \eep analyses it is obtained through a fit to the data it can 
include, besides correlations, also other contributions which are neglected or 
not adequately described by the model. It can be correctly identified with the 
s.f. only if all the theoretical ingredients contributing to the cross sections 
are well under control. On the other hand, the fact that the model, with 
its phenomenological ingredients, is able to give an excellent description of 
\eep data in a wide range of nuclei and in different kinematics~\cite{bof96} gives support and consistency to the interpretation of the s.f. extracted from the comparison between experimental and theoretical cross sections in terms of correlations. 

The source of the depletion has been investigated by using various
methodologies which consider different types of correlations and it was found that the contribution of SRC is small, only a few percent, at most 10-15\%  
when also tensor correlations are included~\cite{mue95,van98,gai00,iva01,roh04,sub08}. The remaining and
larger part of the quenching is due to long-range correlations (LRC)~\cite{dic04,bar09,cip}, collective excitations of the nucleons at the nuclear surface which are related to the coupling between the s.p. motion and collective surface vibrations.
Valence-state protons are emitted at the nuclear surface and it is quite natural that for these states LRC are important. For the removal of protons from deep states, at larger values of $E_{\mathrm m}$,  the role of SRC might be larger. 

The depletion due to SRC is compensated by the admixture of high-momentum components in the s.p. wave function. The first idea would be to investigate SRC in \eep experiments at large values of $p_{\mathrm m}$~\cite{bob,mon}, but calculations of the one-body density matrix and if its diagonal part, the momentum distribution, indicate that the missing strength due to SRC is found not only at large  values of $p_{\mathrm m}$ but also at large values of $E_{\mathrm m}$, in the continuum part of the spectrum. It is therefore interesting to consider the \eep reaction at high missing energies.

\section{The \eep reaction at high missing energies}
\label{sec:miss}

\begin{figure}
\begin{center}
\includegraphics[scale=0.5, angle=0]{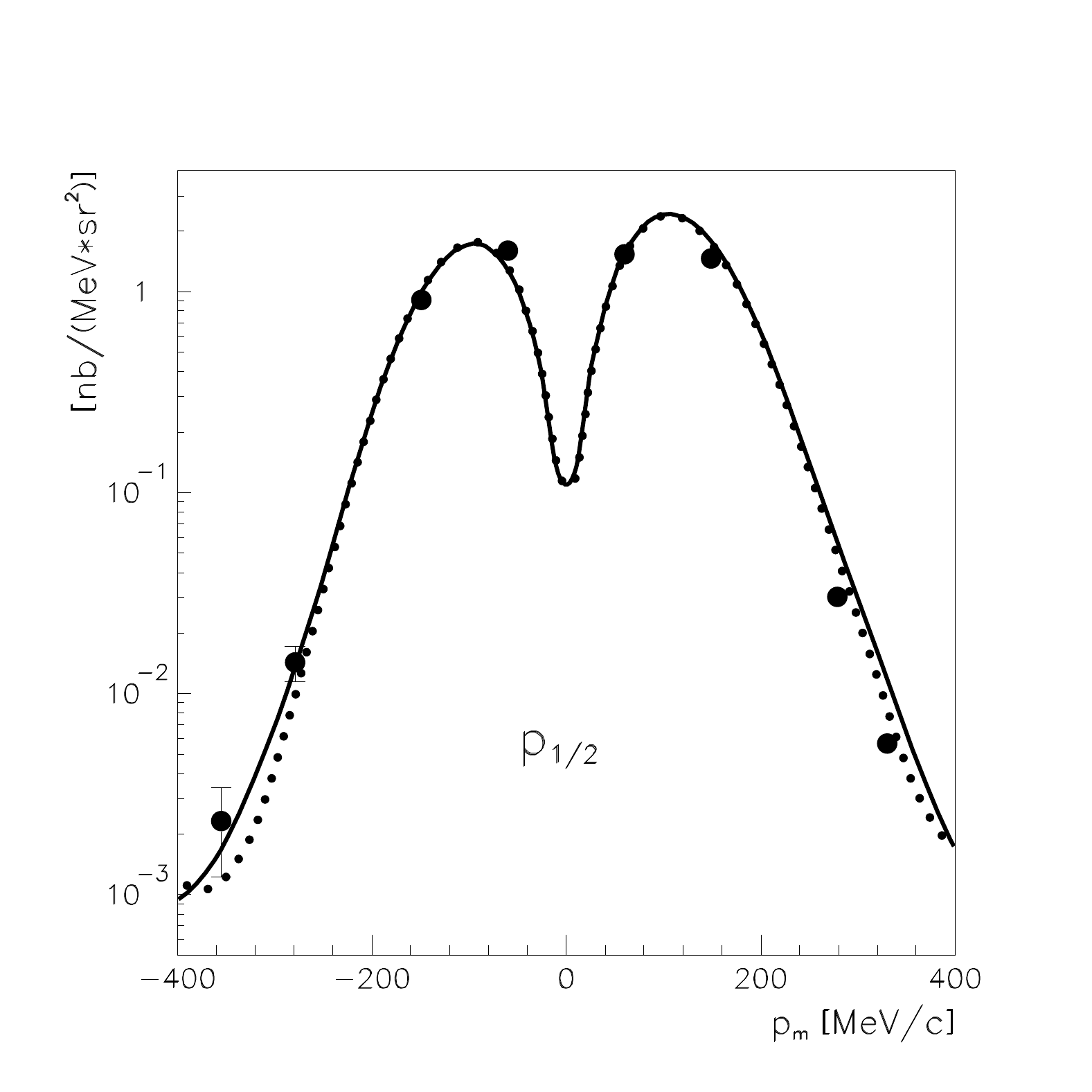} 
\includegraphics[scale=0.5, angle=0]{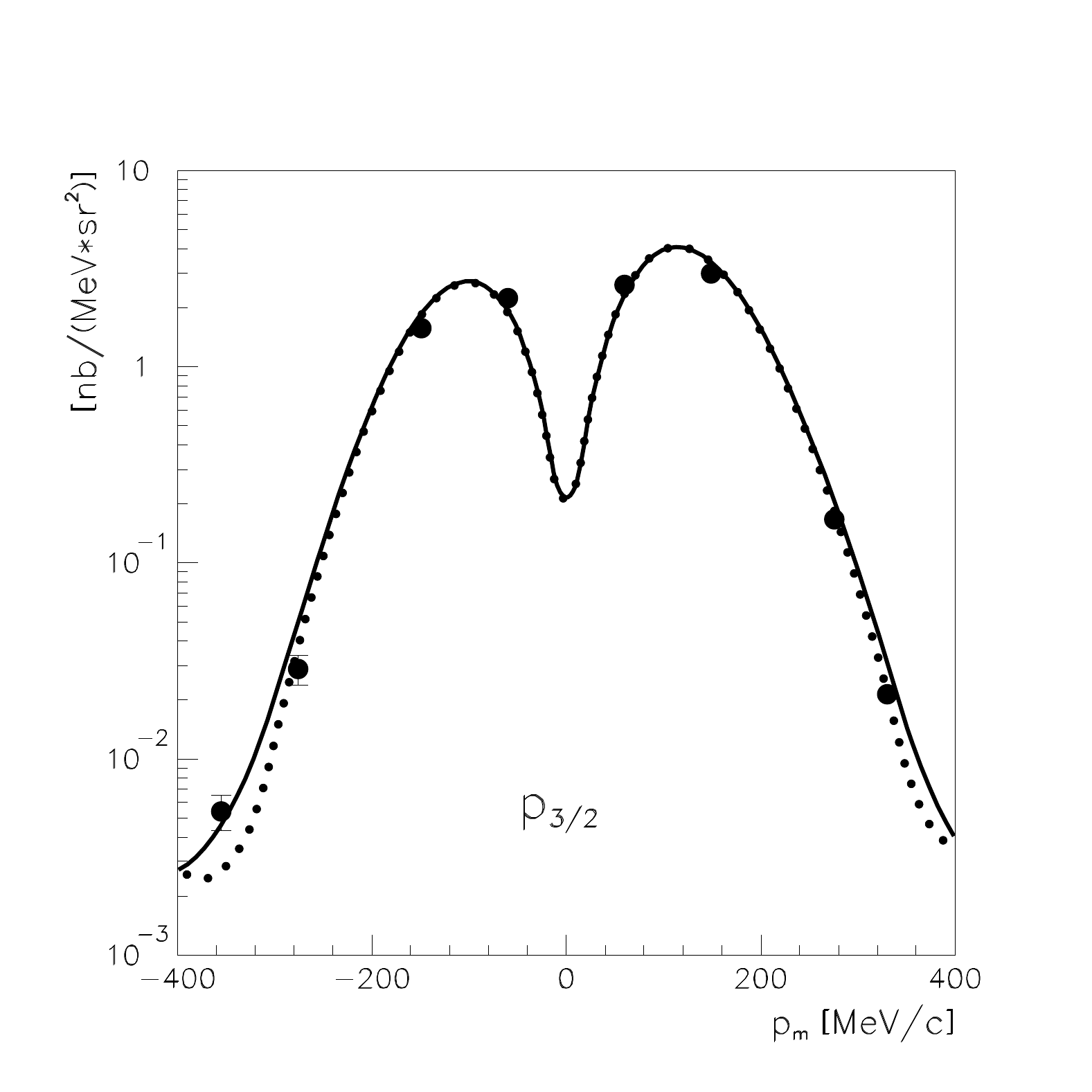} 
\caption{ Cross section of the $^{16}$O\eep reaction for the transitions 
to the $1/2^{-}$ ground state (left panel) and to the $3/2^{-}$ excited state of $^{15}$N in a kinematics with constant ($\q,\omega$), 
with $E_k= 2445$ MeV and an outgoing proton energy of $433$ MeV. The solid and dashed lines give the RDWIA results obtained with two different relativistic OPs. Experimental data from Ref.~\cite{Gao00} (from Ref.~\cite{meu01a}). 
}
\label{fig:rdwia}
\end{center}
\end{figure}

The evolution of the \eep cross section as a function of $E_{\mathrm m}$ and $p_{\mathrm m}$ has been measured at Jefferson Lab for the $^{16}$O\eep reaction~\cite{liy}. The results are shown in Fig.~1 of Ref.~\cite{liy}. 
In the $E_{\mathrm m}$ distribution, for  $E_{\mathrm m} \leq 20$ MeV, there are two sharp peaks, at $E_{\mathrm m}$ = 12 and 18 MeV, that can be attributed to proton knockout from the $1p_{1/2}$ and $1p_{3/2}$ shells. For both states the shape of the experimental cross section as a function of $p_{\mathrm m}$ is well described, up to  
$p_{\mathrm m}$=350 MeV/$c$, by RDWIA calculations~\cite{Gao00,udi99,meu01a}, 
as shown in Fig.~\ref{fig:rdwia}.

For $E_{\mathrm m} > 20$ MeV, in the continuum  part of the spectrum, the experimental cross section exhibits a different behavior~\cite{liy}. At 
$p_{\mathrm m} \approx 50$ MeV/$c$, there is a broad  wide peak centered at $E_{\mathrm m} \approx 40$ MeV, which can be attributed to knockout of protons from the $1s$ state. A RDWIA calculation for proton knockout from the $1s$ state~\cite{kel99} is indeed  able to give a good description of the experimental cross section in the peak region~\cite{liy}. We note that in the continuum the spectral strength has a Lorentzian shape~\cite{bof96,mah} and in the calculations the cross section is spread over $E_{\mathrm m}$ with a Lorentzian parametrization~\cite{liy,mah}. At $p_{\mathrm m} \approx 145$ MeV/$c$ the peak at $E_{\mathrm m} \approx 40$ MeV is less prominent  and for $p_{\mathrm m} \ge 200$ MeV/$c$ it has vanished beneath a flat background.  At $E_{\mathrm m} > 60$ MeV or 
$p_{\rm m} > 200$ MeV/$c$, the experimental cross section does not depend on
$E_{\mathrm m}$ and decreases only weakly with $p_{\mathrm m}$. 
Increasing $p_{\mathrm m}$, the agreement between RDWIA results and data gets worse and proton knockout from the $1s$ state accounts for only a small or even negligible fraction of the experimental cross section. The contribution of $(e,e^{\prime}pn)$ and $(e,e^{\prime}pp)$ knockout to \eep~\cite{ryc}  increases increasing $p_{\mathrm m}$ and gives a flat cross section, consistent with the data, but also this contribution accounts for only a part of the measured cross section. This means that increasing  $E_{\mathrm m}$ and $p_{\mathrm m}$ the \eep cross section includes other contributions beyond one- and two-nucleon knockout.
Among other possible contributions the role of FSI is considered and discussed in the following. 

For the exclusive \eep reaction FSI are usually described in DWIA and RDWIA by a complex OP where the imaginary part gives an absorption that 
reduces the calculated cross section. Such a reduction is essential to reproduce \eep data for the removal of valence protons. The description of FSI in terms of an absorptive OP may be not suited at high missing energies and missing momenta. 

The imaginary part of the OP accounts for the fact that in elastic nucleon-nucleus scattering, 
if other channels are open besides the elastic one, part of the incident flux is lost in the elastically scattered beam and goes to the inelastic channels that are open. In the DWIA the imaginary part removes the contribution of inelastic channels. This approach can be correct for the exclusive \eep reaction at low values of $E_{\mathrm m}$, where a discrete final state is selected and it is reasonable to assume that the experimental cross section receives contributions mainly from the one-nucleon knockout process where  the outgoing nucleon scatters elastically with the residual nucleus in the selected final state. In contrast, the DWIA would be wrong for the inclusive scattering, where all the final-state channels are included, the flux lost in a channel must be recovered in the other channels, and in the sum over all the channels the flux can be redistributed but must be conserved. In every channel flux is lost towards other channels and flux is gained due to the flux lost in the other channels  just toward that channel. The DWIA accounts only for the flux lost and not for the flux gained and therefore does not conserve the flux.

A different but consistent model to account for FSI  has been  developed for the inclusive scattering, the so-called Green's function (GF) model~\cite{eenr,eesym,ee}. Under many aspects the model is still based on the IA, i.e. on the assumption that the probe interacts through a one-body current only with an ejectile nucleon, but in the inclusive scattering a sum is performed over all the nucleons of the target. Consistently with the DWIA used for the exclusive scattering, FSI are accounted for in the GF model by the same complex OP, but the GF formalism translates the flux lost toward inelastic channels, represented by the imaginary part of the OP, into the strength observed in the inclusive reaction. In the model, with suitable approximations, the components of the inclusive response are obtained in a form where the basic ingredients are the same matrix elements as in  Eq.~(\ref{eq:dwia}), but in the scattering states there  are eigenfunctions of the OP and of ist Hermitian conjugate, where the imaginary part has a different sign: one sign gives an absorption, a loss of flux, the opposite sign gives a gain of flux. Thus, in the model the imaginary part redistributes the flux over all the final-state channels and in the sum the total flux is conserved. 

In the GF model the OP becomes a powerful tool to include, in a simpler and somewhat less model dependent way than with microscopic  calculations, important inelastic contributions beyond one-nucleon knockout (such as, for instance, multi-nucleon processes, rescattering, some non nucleonic contributions) which are not included in usual descriptions of FSI based on the IA.
The GF model was originally developed within a nonrelativistic~\cite{eenr,eesym} and then within a relativistic framework~\cite{ee,eea,confee,ex} for the inclusive \ee scattering. The relativistic GF (RGF) model has been extended to neutrino-nucleus scattering~\cite{cc,acta,acta1,confcc,compmini,prd,prd1,prd2,compnc,prd3,prd4,martin,GRFOP}. 
The model has been quite successful in the comparison with QE \ee data, both for electron and neutrino-nucleus scattering.  

For the reaction \eep in the continuum, at high values of 
$E_{\mathrm m}$ and $p_{\mathrm m}$ one-nucleon knockout gives only a part of the cross sections, other processes and reaction mechanisms become important, and  we can expect that the cross section contains some of the inelastic contributions which are removed in DWIA by the imaginary part of the OP and which are recovered by the imaginary part of the OP in the GF model. The use of phenomenological OPs in DWIA and GF calculations does not allow us to disentangle and evaluate the role of a specific inelastic process, but an important contribution can be expected  from rescattering, multiple scattering of the ejected proton with the residual nucleus, i.e. the process where the probe interacts with the ejectile proton which, in the continuum, is removed from a deep state, from the nuclear interior, and in its way through the nucleus subsequently undergoes a series of secondary collisions with other nucleons where it can change direction and lose energy before being emitted and detected. These contributions are removed in DWIA by the OP but can be included in the \eep cross section at high missing enrgies. 

A quantitative estimate of the multi-scattering effects in \eep at high $E_{\mathrm m}$ has been given in Refs.~\cite{viv1,viv2,viv3} in order to improve the usual treatment of FSI	by means of an OP.
The method follows the lines of the multi-step direct (MSD) scattering theory of Feshbach, Kerman, and Koonin~\cite{fkk80,msd}, which allows one to trace the secondary collisions of the emitted nucleon. The proton, following the initial electromagnetic interaction, is excited to the continuum and subsequently undergoes a series of two-body interactions with the residual nucleus before being emitted. Thus, there are a series of collisions leading to intermediate states  of increasing complexity. At each step a nucleon can be emitted. The theory combines a quantum-mechanical treatment of multiple scattering with statistical assumptions that lead to the convolution nature of the multistep cross section and enables the calculation of higher-order contributions (up to six steps) which would be otherwise impracticable.

The cross section for a scattered electron of energy $E_{k'}$ and angle $\Omega_{k'}$ and ejectile proton of energy $E$ and angle $\Omega$ is written as an incoherent sum of a one-step and multistep ($n$-step) terms
\beq
\frac{{\rm d}^{4}\sigma}{{\rm d}\Omega_{k'} {\rm d}E_{k'} {\rm d}\Omega {\rm d}E} 
= 
\frac{{\rm d}^{4}\sigma^{(1)}}{ {\rm d}\Omega_{k'} {\rm d}E_{k'} {\rm d}\Omega
{\rm d}E } 
+ \sum_{n=2}^{\infty}\frac{{\rm d}^{4}\sigma^{(n)}}{ {\rm d}\Omega_{k'}
{\rm d}E_{k'} {\rm d}\Omega {\rm d}E}.
\label{eq:msdeep}
\eeq

The one-step term corresponds to one-nucleon knockout and is calculated within the DWIA in the continuum. Th $n$-step term is given by a convolution of the direct \eep knockout cross sections and one-step MSD cross sections over all intermediate
energies $E_{1},\, E_{2}\dots$ and angles $\Omega_{1},\,\Omega_{2}\dots$ obeying
energy and momentum conservation rules:
\begin{eqnarray}
\frac{{\rm d}^{4}\sigma^{(n)}}{ 
{\rm d}\Omega_{k'}{\rm d}E_{k'} {\rm d}\Omega {\rm d}E} 
& =&  \left(\frac{m}{4\pi^{2}}\right)^{n-1} 
\int {\rm d}\Omega_{n-1}\int {\rm d}E_{n-1}E_{n-1}\dots \nonumber \\
& & \times \int {\rm d}\Omega_{1}\int {\rm d}E_{1}E_{1} 
\frac{{\rm d}^{2}\sigma^{(1)}}{{\rm d}\Omega dE}(E,\Omega
 \leftarrow E_{n-1},\Omega_{n-1})\dots \nonumber \\
& &\times \frac{{\rm d}^{2}\sigma^{(1)}}{{\rm d}\Omega_{2}
 {\rm d}E_{2}}(E_{2},\Omega_{2} \leftarrow E_{1},\Omega_{1}) 
\frac{{\rm d}^{4}\sigma}{{\rm d}\Omega_{k'}{\rm d}E_{k'} {\rm d}\Omega_{1} dE_{1}}.
\label{eq:msdneep}\\ \nonumber
\end{eqnarray}
The one-step MSD cross sections for the subsequent NN
scatterings are calculated by extending the DWBA theory to the continuum~\cite{viv1}.

The method follows the trace of the fast emitted proton but does not follow the fate of secondary nucleons, which become more energetic increasing $E_{\mathrm m}$. 
It does not include either two-nucleon knockout (that in particular kinematic situations can be important) or pion production and absorption, which can be very important and even dominant increasing the energy. As a consequence, the method is suited to study the continuum  but for not too high values of $E_{\mathrm m}$. The neglected processes might be included in the model, but this would be technically quite involved. 

\begin{figure}
\begin{center}
\includegraphics[scale=0.5, angle=0]{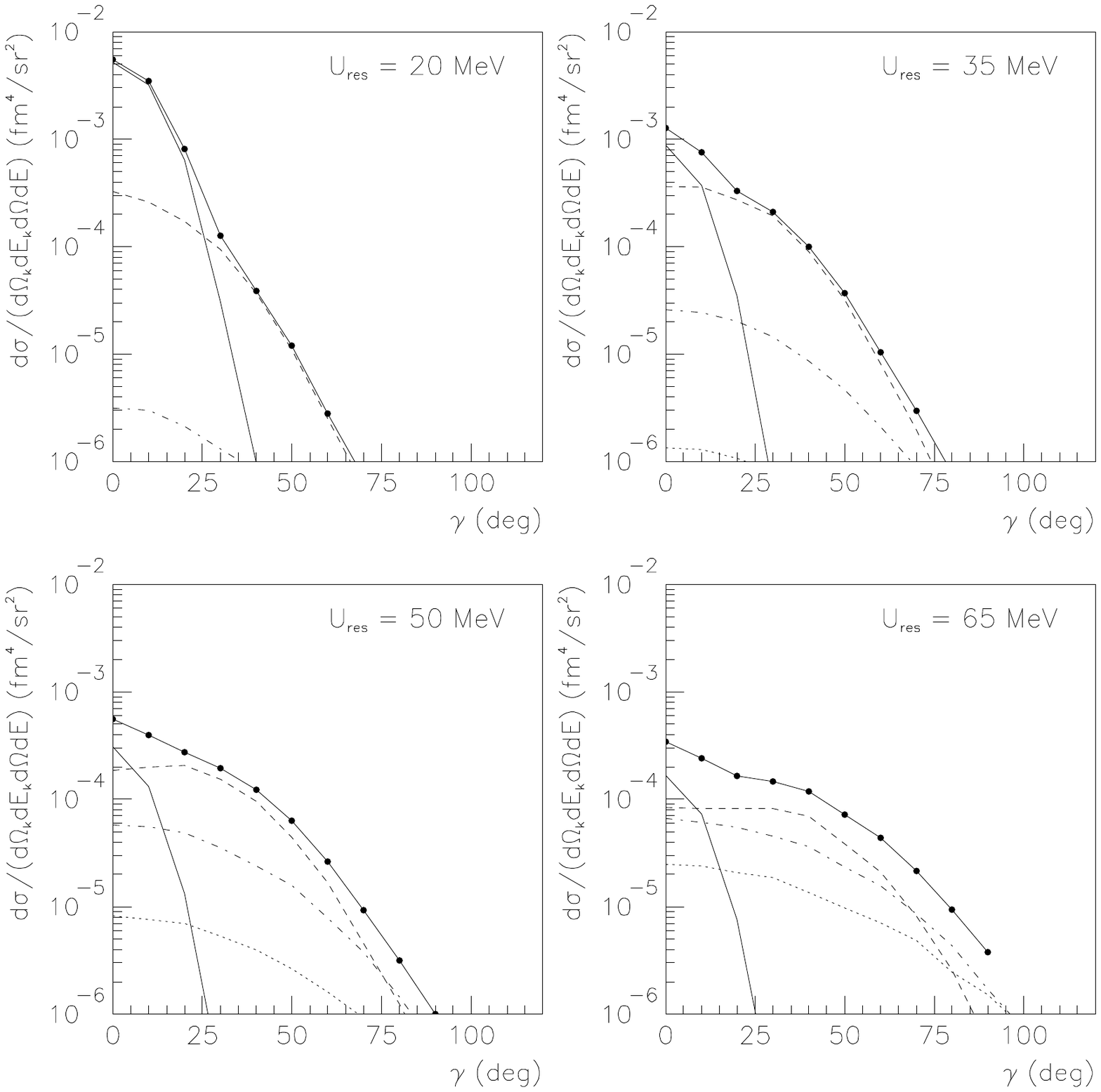} 
\caption{Cross section of the $^{40}$Ca\eep reaction as a function of
the scattering angle $\gamma$ between $\p'$ and $\q$ at different energies of the residual nucleus $U_{\mathrm res}= E_{\mathrm m} - E_{\mathrm b.e.}$, $E_{\rm b.e.}$ is the g.s. nucleon binding energy (8 MeV). Calculations are performed in a kinematics with $(\q, \omega)$ constant where $E_k= 497$ MeV, $E_{k'}= 350$ MeV, the electron scattering angle $52.9^{\circ}$, and the outgoing proton energy $87\pm 10$ MeV. Solid line for the direct
\eep process; dashed, dot-dashed, and dotted lines for the two-, three-, and four-step
processes. The total result is given by the solid line with marking dots. 
(from Ref.~\cite{viv1}). 
}
\label{fig:resc}
\end{center}
\end{figure}
An example is shown in Fig.~\ref{fig:resc}, where the direct \eep knockout and multistep cross sections are displayed for the $^{40}$Ca\eep reaction at four increasing values of $E_{\mathrm m}$.  At the lowest excitation energies, the direct nucleon knockout process dominates and  exhibits a strong forward-peaking. The multistep contributions are important at large  scattering angles, corresponding to larger values of $p_{\mathrm m}$, over the whole energy
range. We note that $p_{\mathrm m}$ increases increasing the missing energy and the scattering angle. With increasing excitation energy the two-step and three-step contributions
become gradually more important than the one-step direct process over most of the angular range, apart from the very small angles $\leq
10^{\circ}$. The domination of multistep  processes at large scattering angles is expected since, as a result of multistep scattering, the leading proton gradually loses memory of its initial direction yielding thus increasingly symmetric angular distributions. 

This example confirms that increasing $E_{\mathrm m}$ and $p_{\mathrm m}$ other mechanism beyond direct one-nucleon knockout become gradually more important and even dominant.
Although based on a full quantum-mechanical treatment of multiple scattering, the method contains approximations that restrict its applicability to a limited range of missing energies. At larger $E_{\mathrm m}$ pion production and nucleon resonances contributions should be included.  
Alternative methods of dealing with FSI, which, although based on some semi-classical assumptions, can account for more contributions can be found, for instance, in Refs.~\cite{val,gibu}. 

In the \eep reaction processes and mechanism beyond direct one-nucleon knockout become increasingly important by increasing $E_{\mathrm m}$ and $p_{\mathrm m}$. The role and relevance of a specific contribution is very sensitive 
to the kinematic conditions. The flexibility of electron scattering to independently vary the energy and momentum transfer, and therefore to explore different kinematic regions, makes it possible to envisage and select suitable kinematics for the experiments able to minimize or emphasize and therefore investigate specific contributions.  Electron scattering experiments can be selective. In neutrino experiments it is much more difficult to be selective.

\section{From electron scattering to neutrino scattering}
\label{sec:neutrino}

Models developed for electron scattering have been extended to describe nuclear effects in neutrino-nucleus scattering. The extension of the formalism is straightforward, the main difference being between the electromagnetic and weak currents. Schematically, the nuclear response as a function of the energy and momentum transfer is the same for electron and neutrino scattering. Different kinematic regions, characterized by different reaction mechanisms, occur, in particular the QE peak, which is  dominated by the s.p. dynamics and by the process of one-nucleon knockout.

In charged-current QE (CCQE) scattering one nucleon is emitted, i.e.
\begin{equation}
\nu (\bar\nu) + \mathrm{A}  \rightarrow  l^{-} (l^{+}) +
\mathrm{p}(\mathrm{n}) + ( \mathrm{A}-1). 
\end{equation}
The situation where only the final lepton is detected can be treated with same models developed for the inclusive QE \ee scattering, such as, for instance, with the RGF model. 
From the experimental point of view, however, there is a fundamental difference between electron and neutrino scattering. In electron scattering experiments the incident electron energy is basically known and $\omega$ and $q$ are clearly determined. In contrast, in (anti)neutrino scattering experiments, the incident (anti)neutrino energy is not known. It can be reconstructed on the basis of a model, which, however, requires the assumption of a specific reaction mechanism. A good determination of the neutrino energy is very important for the analysis of neutrino oscillation experiments, but the determination requires a good control of the reaction mechanism and of all nuclear effects. Anyhow, since experimentally the incident energy is not known, $\omega$ and $q$ are not determined, and the measured cross sections are usually averaged over the neutrino flux. 
Calculations for the comparison with data must be performed for all the values of the incident energy for which the (anti)neutrino flux has considerable strength, therefore in different kinematic situations. The flux-average procedure picks up contributions from different kinematic regions, with different values of $\omega$, not only from the QE region (here QE means that no pions are detected in the final state, i.e. CCQE means CC$0\pi$) and contributions other than one-nucleon knockout can be more important than in the inclusive QE \ee scattering. 

The RGF results for the CCQE scattering where only the final lepton is detected indicate that the relevance of the inelastic contributions beyond one-nucleon knockout recovered by the imaginary part of the OP is greater than in the inclusive \ee scattering. The model has been quite successful in the description of CCQE MiniBooNE and MINER$\nu$A data, both for neutrino and antineutrino scattering~\cite{miniboone,minibooneanti,compmini,prd1,minerva,minerva1,prd4}, and of neutral-current elastic (NCE) MiniBooNE data~\cite{prd,prd3,mbnce,mbnceanti}.
\begin{figure}
\begin{center}
\includegraphics[scale=0.3, angle=0]{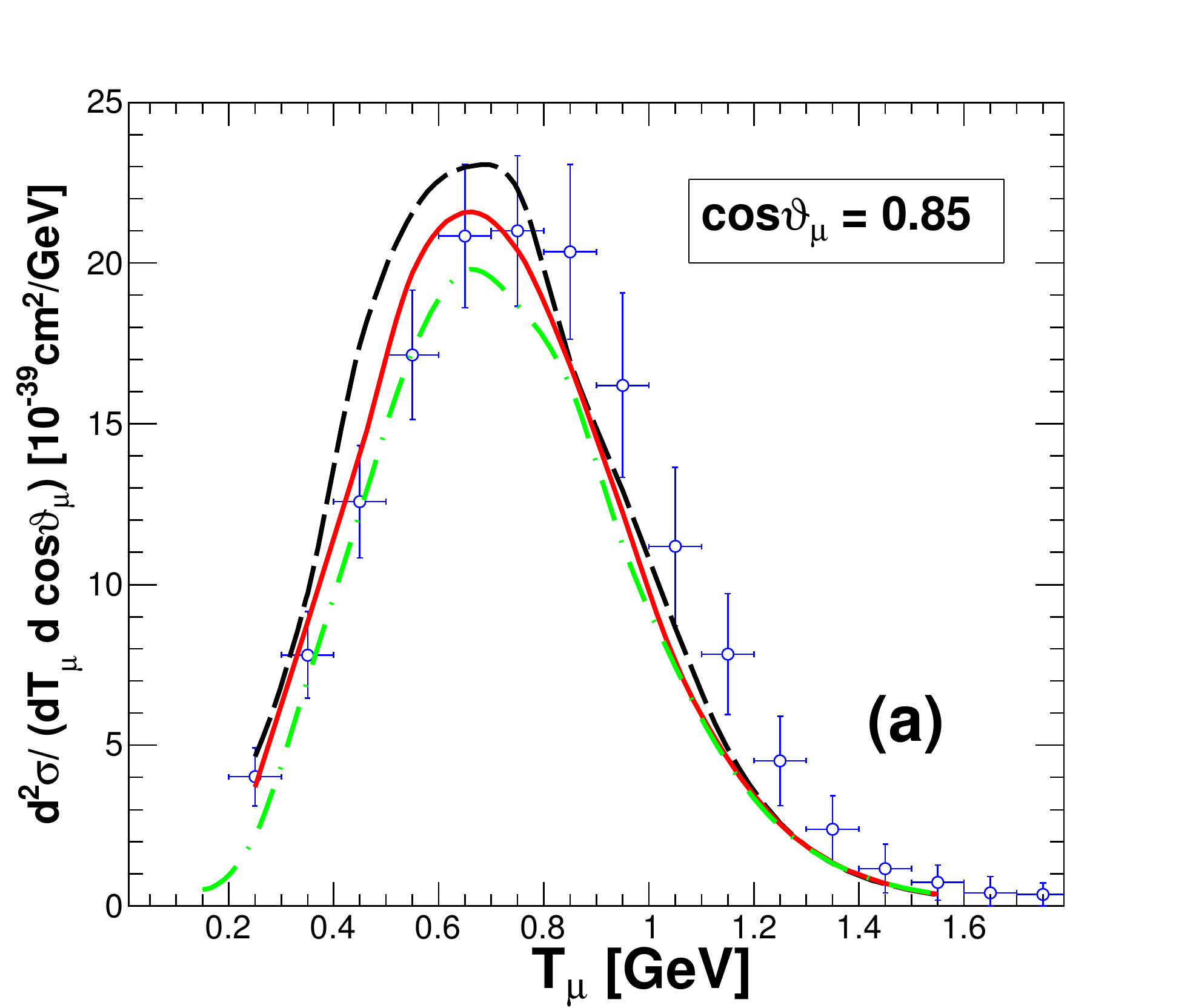} 
\includegraphics[scale=0.3, angle=0]{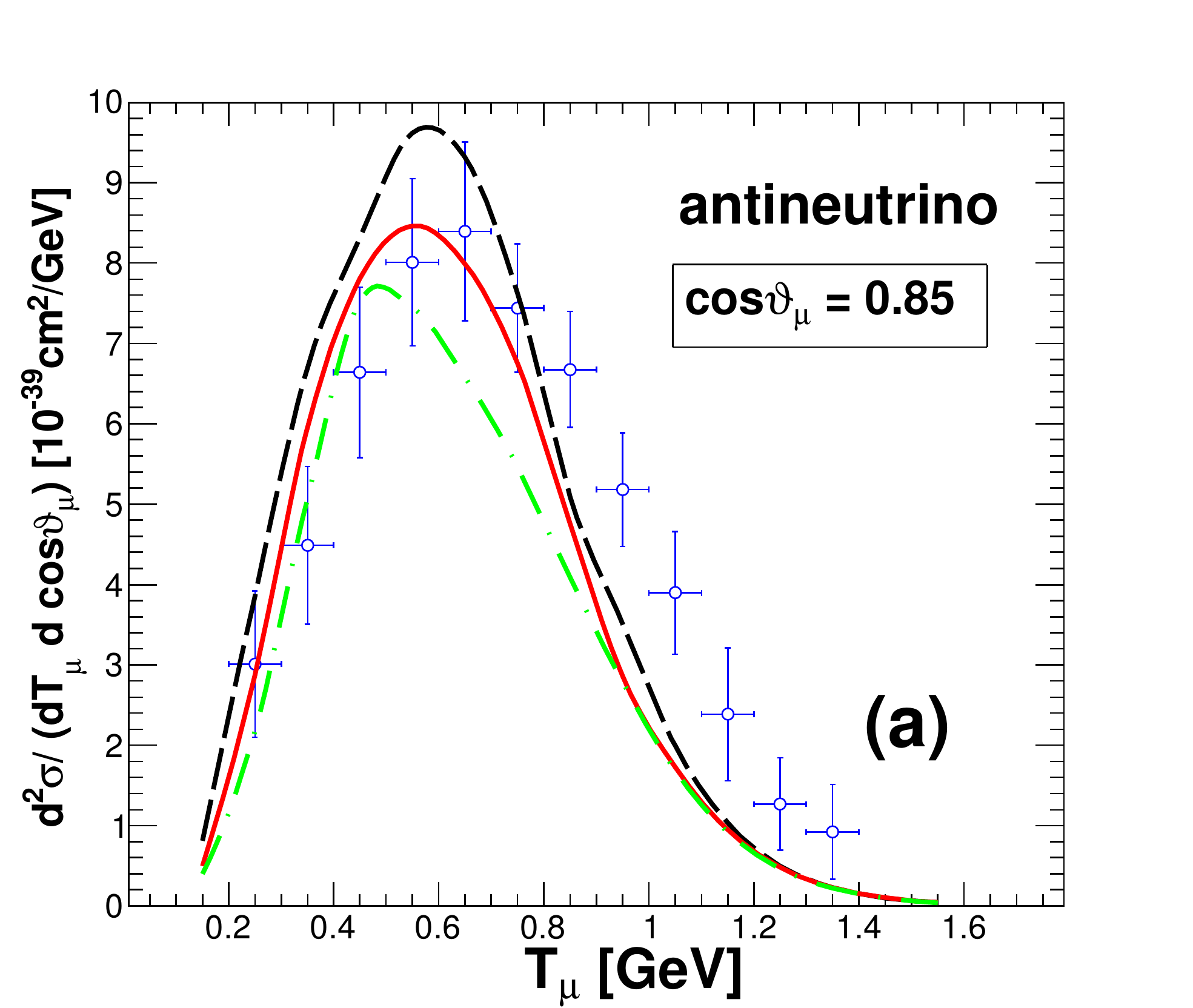}\\ 
\includegraphics[scale=0.3, angle=0]{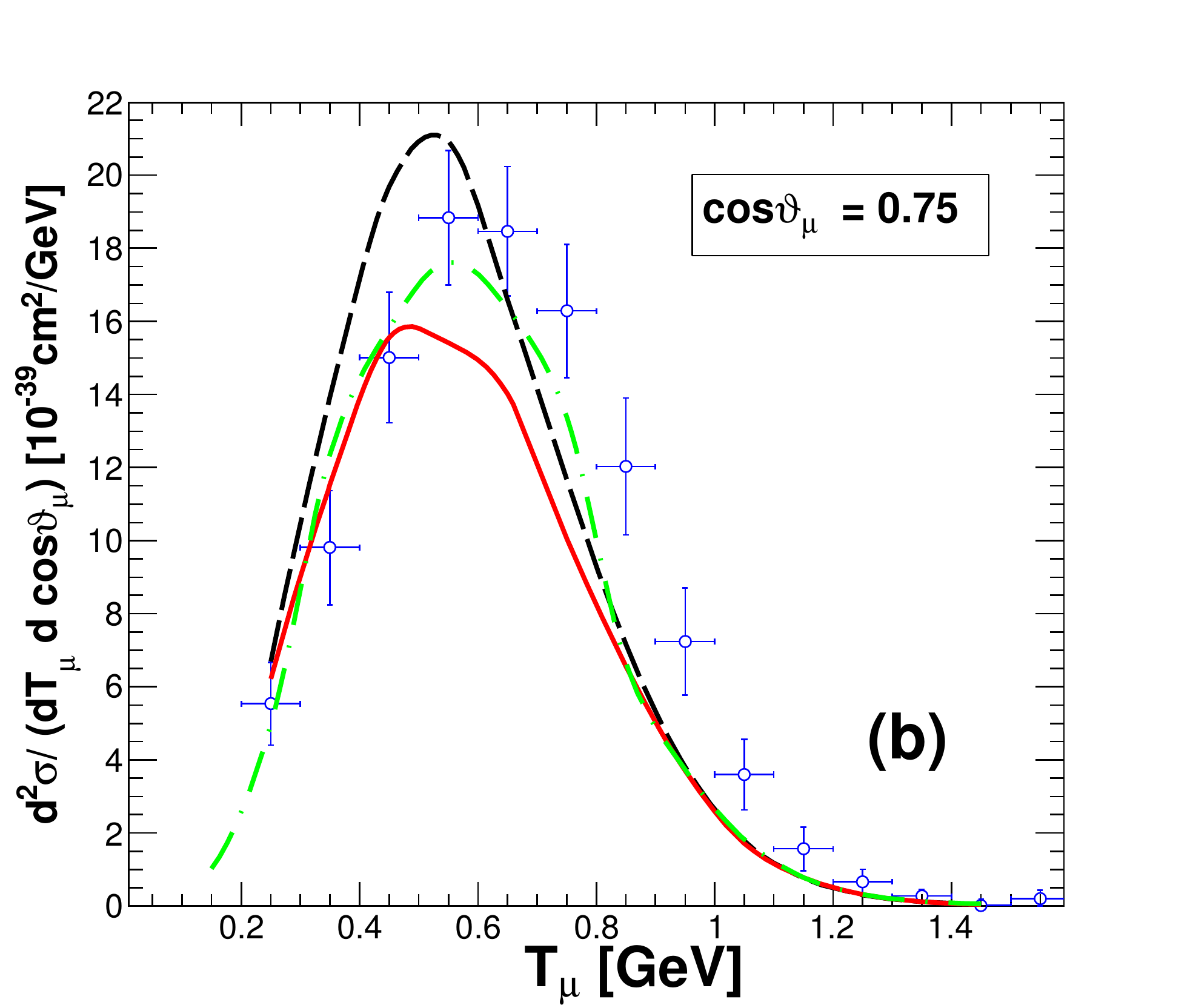} 
\includegraphics[scale=0.3, angle=0]{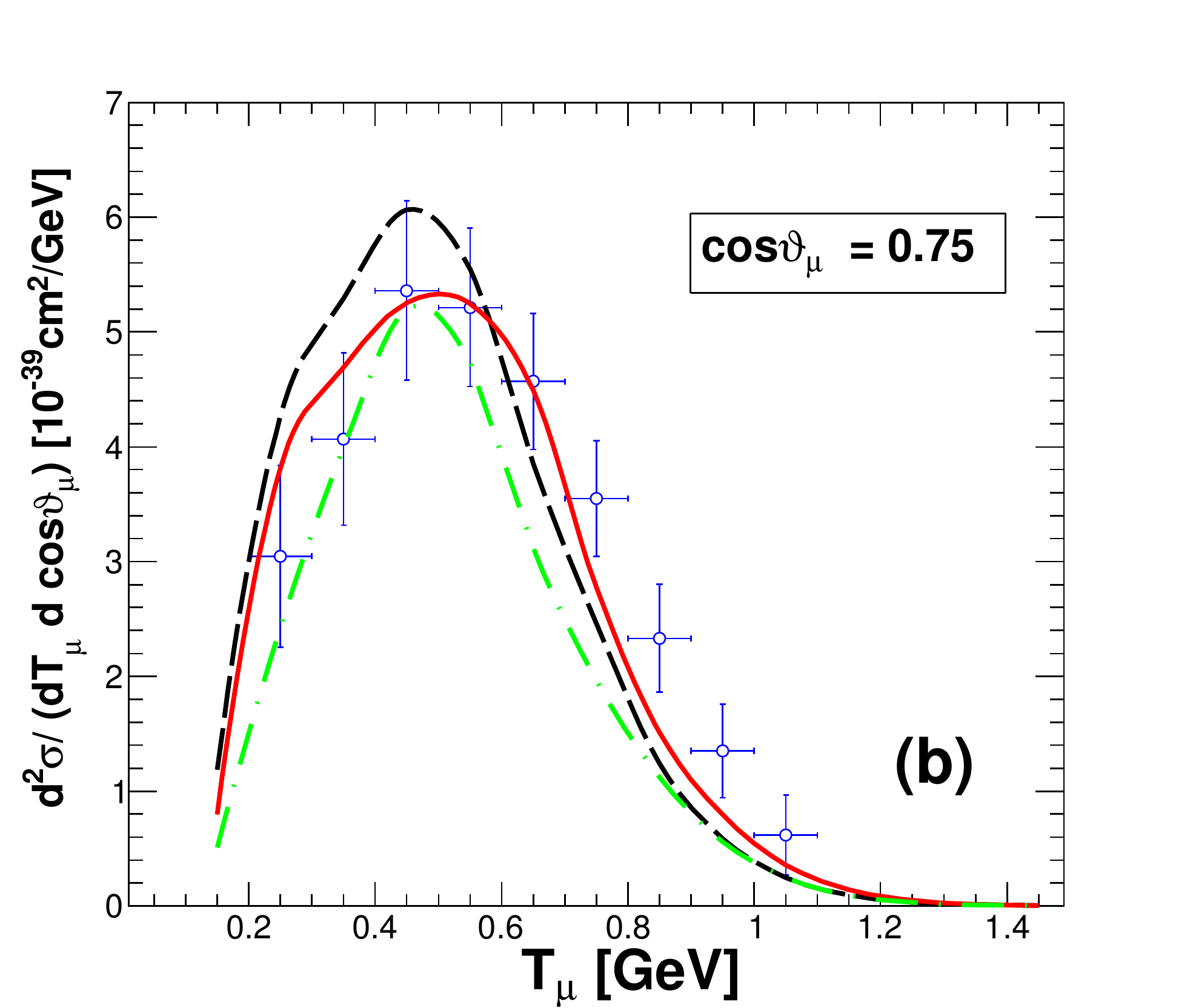}\\ 
\includegraphics[scale=0.3, angle=0]{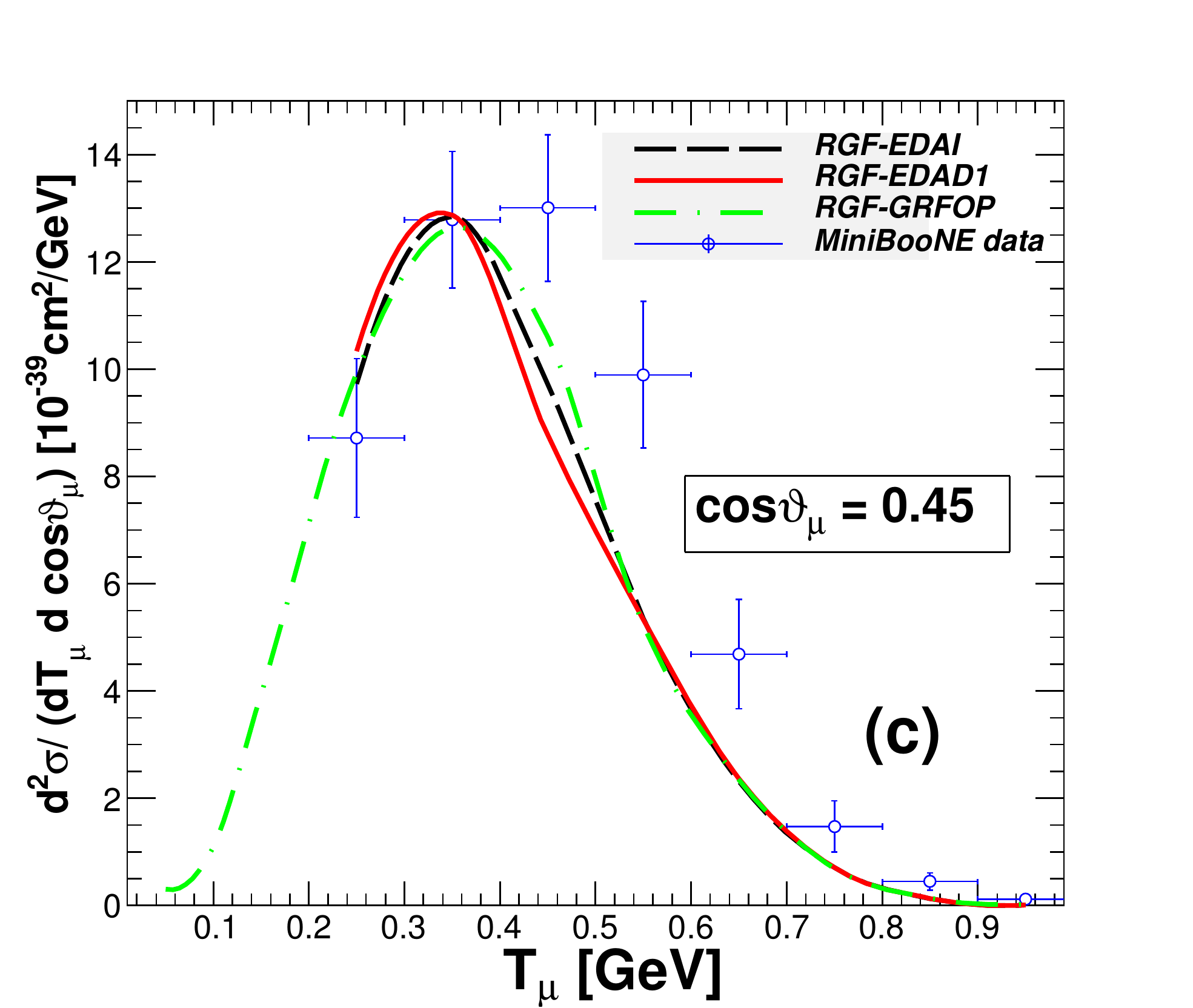} 
\includegraphics[scale=0.3, angle=0]{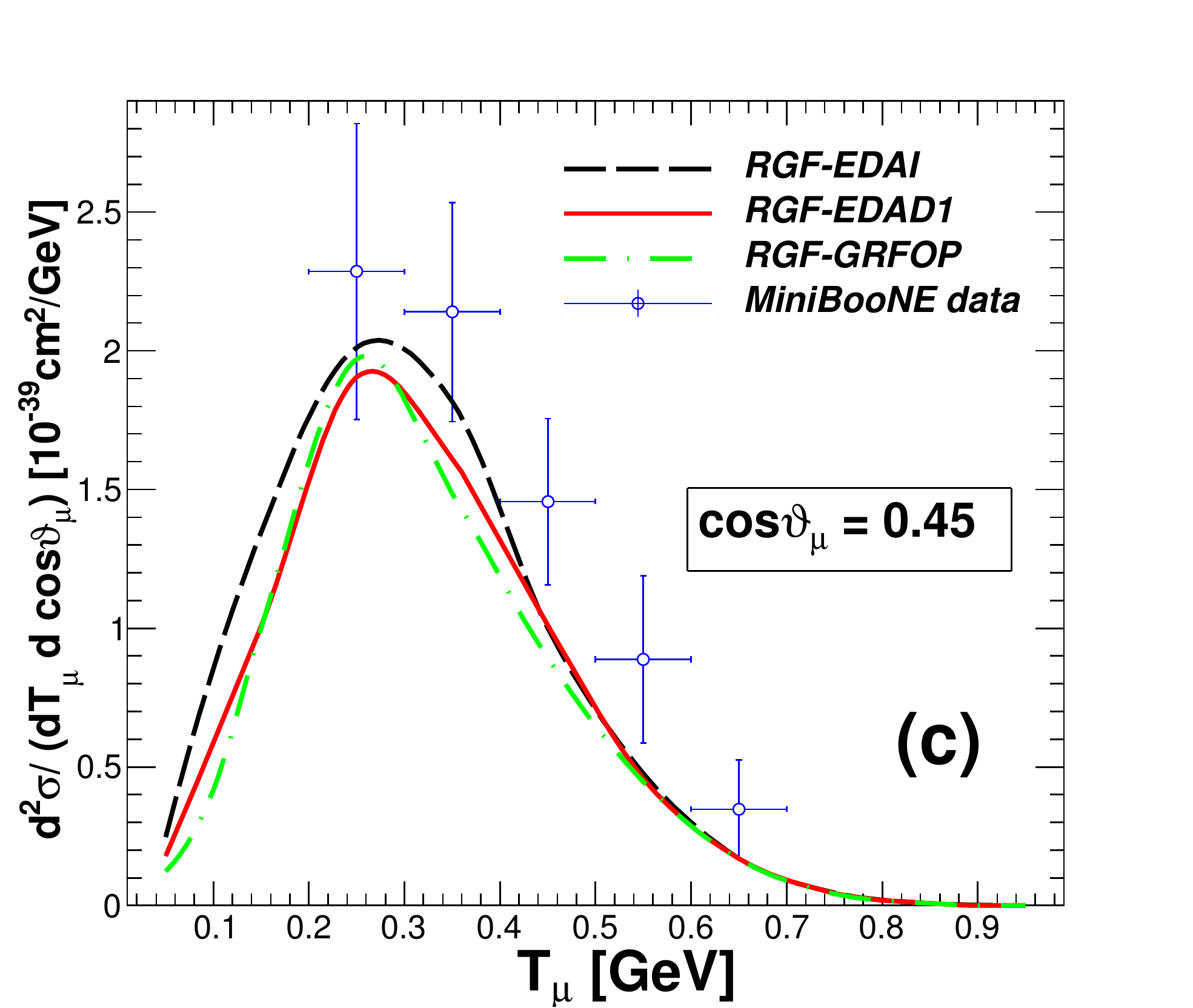} 
\caption{Flux-averaged double-differential cross section per target nucleon for the CCQE $^{12}$C$(\nu_{\mu} , \mu^-)$ (left panels) and 
$^{12}$C$(\bar\nu_{\mu} , \mu^-)$ (right panels) reactions as a function of the muon kinetic energy $T_{\mu}$ for three bins of the muon scattering angle $\cos\vartheta_\mu$. The RGF results obtained with three different relativistic OPs are compared. Experimental data from MiniBooNE \cite{miniboone, minibooneanti}. (from Ref.~\cite{GRFOP}).}
\label{fig:ddneu}
\end{center}
\end{figure}
An example is shown in Fig.~\ref{fig:ddneu}, where the CCQE MiniBooNE cross sections are compared with the RGF results obtained with three different relativistic OPs. The OP is the crucial ingredient of the RGF model, different OPs are available for the calculations, which basically differ in the imaginary part. The different imaginary parts do not affect significantly elastic observables, but can give different inelastic contributions and therefore different results when the OPs are employed in RGF calculations.  

If also the emitted nucleon is detected, the situation is similar to the \eep reaction and in principle the same models can be applied. However, since the incident neutrino energy is not known, calculations are required in a  
range of energies where the neutrino flux has considerable strength, corresponding to different values of $E_{\mathrm m}$, both in the discrete and in the continuum part of the spectrum, where various effects and reaction processes involving more than one nucleon may come into play. A model is required able to include, as consistently as possible,  one-nucleon knockout, FSI, two-nucleon knockout, mesons and deltas, meson-exchange currents etc. 
This is a challenging task.   

\section{Conclusions}
\label{sec:concl}

The work done over several decades on electron scattering, which has provided a wealth of detailed information on nuclear properties, is extremely useful also for the analysis and the interpretation of data from neutrino experiments.

 
The comparison with electron scattering data is the first necessary test of the validity and of the predictive power of a nuclear model. Any model to be applied to neutrino-nucleus scattering must, as a first step, be tested in comparison with electron scattering data.

The experimental neutrino-nucleus  cross sections contain contributions from different kinematic regions, where different effects and reaction mechanisms may come into play. A reliable interpretation of data requires that all these  contributions are taken into account and evaluated as accurately as possible. 
To this aim, electron scattering experiments in suitable kinematics can be helpful to disentangle and separately investigate specific effects and reaction mechanisms. 

The information that can be obtained from new electron scattering experiments focussed on specific issues, besides being interesting in its own right, can be exploited for the analysis of neutrino experiments. 
For instance, as  future  neutrino  experiments will  use  large  liquid  argon  detectors, the dedicated $^{40}$Ar\eep experiment at Jefferson Lab~\cite{ar,VP} will provide the experimental input indispensable to construct the argon spectral function, thus paving the way for a reliable estimate of the neutrino cross sections.




\bibliographystyle{elsart-num} 

\end{document}